\documentclass[]{JHEP3}
\usepackage{epsfig}
\usepackage{amsmath}
\def\be{\begin{equation}}
\def\ee{\end{equation}}
\def\bea{\begin{array}}
\def\beqa{\begin{eqnarray}}
\def\eeqa{\end{eqnarray}}
\def\beqas{\begin{eqnarray*}}
\def\eeqas{\end{eqnarray*}}

\def\bp{\begin{picture}}
\def\ep{\end{picture}}
\def\bc{\begin{center}}
\def\ec{\end{center}}
\def\bfig{\begin{figure}}
\def\efig{\end{figure}}

\def\bit{\begin{itemize}}
\def\eit{\end{itemize}}
\def\nn{\nonumber}
\def\f{\frac}

\def\[{\left[}
\def\]{\right]}
\def\({\left(}
\def\){\right)}

\def\..{\left.}
\def\.{\right.}
\def\tl{\tilde}

\def\ep{\epsilon}

\title{Low energy supersymmetry confronted with current experiments: an overview}

\date{\today}

\author{Fei Wang$^1$\thanks{feiwang@zzu.edu.cn},
Wenyu Wang$^2$\thanks{wywang@bjut.edu.cn}, 
Jin Min Yang$^{3,4}$\thanks{jmyang@itp.ac.cn }, 
Yang Zhang$^1$\thanks{zhangyangphy@zzu.edu.cn}, 
Bin Zhu$^{5}$\thanks{zhubin@mail.nankai.edu.cn}\\

$^1$ School of Physics, Zhengzhou University, Zhengzhou 450000, P. R. China \\
$^2$ School of Science, Beijing University of Technology, Beijing 100124, P. R. China\\
$^3$ CAS Key Laboratory of Theoretical Physics, Institute of Theoretical Physics,
                Chinese Academy of Sciences, Beijing 100080, P. R. China \\
$^4$ School of Physical Sciences, University of Chinese Academy of Sciences, 
                Beijing 100049, P. R. China \\
$^5$ Department of Physics, Yantai University, Yantai 264005, P. R. China 
}

\abstract{This study provides a brief overview of low-energy supersymmetry (SUSY) in light of current experimental constraints, such as collider searches, dark matter searches, and muon $g-2$ measurements. In addition, we survey a variety of low energy supersymmetric models: the phenomenological minimal supersymmetric model (MSSM); the supersymmetric models with cut-off-scale boundary conditions, i.e., the minimal supergravity (mSUGRA) or the constrained MSSM (CMSSM), 
the gauge mediation of SUSY breaking (GMSB), and the anomaly mediation of SUSY breaking (AMSB), 
as well as their extensions. The conclusion is that the low energy SUSY can survive all current 
experimental constraints and remains compelling, albeit suffering from a little fine-tuning problem. The fancy models like mSUGRA, GMSB, and AMSB need to be extended if the muon $g-2$ anomaly comes from new physics.  }   

\begin{document}
\maketitle 
\section{Introduction}

Despite its remarkable phenomenological success, the Standard Model (SM) in particle physics still has remaining puzzles, such as the origin of the free parameters, the matter-antimatter asymmetry, the instability of the electroweak scale, or the divergent quantum correction of Higgs boson mass, and the nature of cold dark matter. Searching for new physics beyond the SM is the central theme of today and future particle physics. The low energy supersymmetry (SUSY) has been the most appealing framework among various new physics hypotheses. Phenomenologically speaking, the SUSY extension of the SM could, solve the hierarchy problem, realize gauge couplings unification, adopt the proper baryogenesis mechanisms, and generate the cold dark matter candidate. 

\begin{figure}[htb]
\begin{center}
\includegraphics[width=7.5cm,height=9cm]{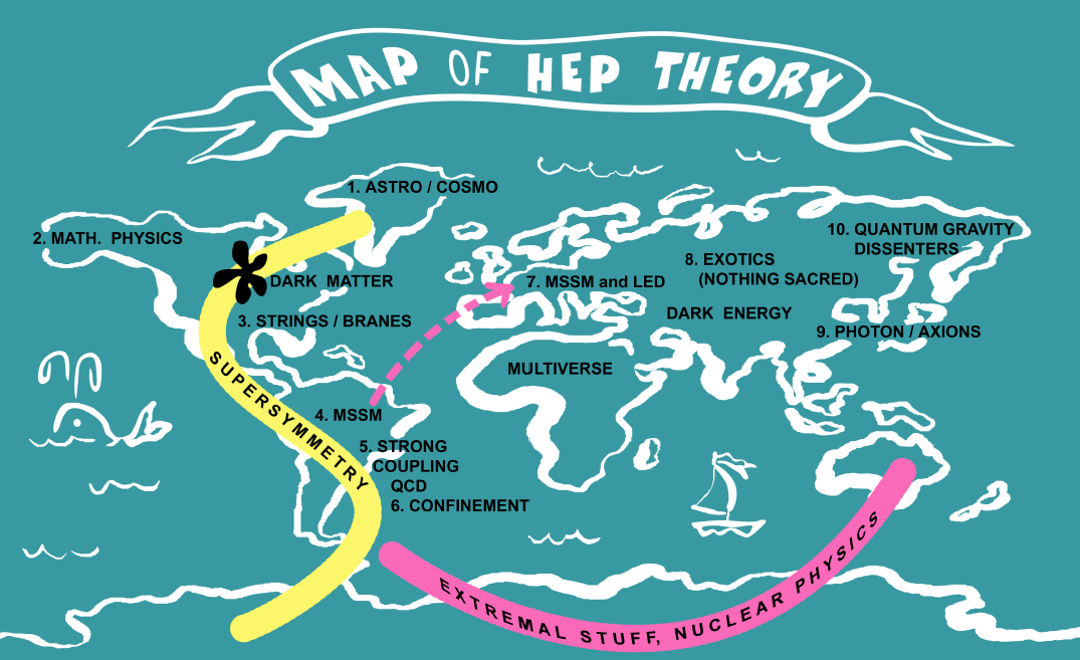}
\includegraphics[width=7.5cm,height=9cm]{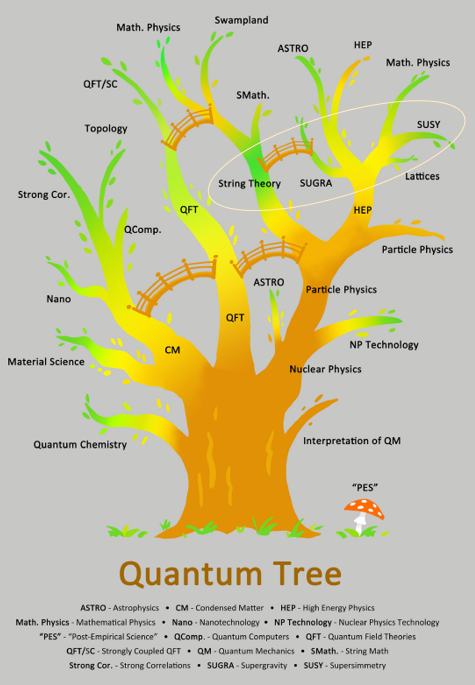}
\end{center}
\vspace{-.5cm}
\caption{The map of high energy physics and the quantum tree taken from \cite{shifman}.}
\label{fig1}
\end{figure}

Most notably, SUSY predicts a neutral CP-even Higgs boson upper bounded roughly 
by 135 GeV,  corroborated by the Large Hadron Collider (LHC) discovery of a 125 GeV Higgs boson.     
From the theoretical view, as a new fundamental symmetry, SUSY is mathematically charming and, 
unlike other miscellaneous new physics models, {\it SUSY is part of a larger vision of physics, 
not just a technical solution \cite{witten}}. 
SUSY has been in the mainstream of high energy physics for almost half a century.
 As depicted in Fig.\ref{fig1} \cite{shifman},  SUSY plays a crucial role in the map of high 
energy physics and even the whole tree of the quantum theory. 
 Indeed, SUSY is needed and predicted by string theory, i.e.,   
 {\it the concept of supersymmetry emerged historically, at least in part because of its role in 
string theory \cite{witten}} .
So far, string theory is the most hopeful candidate for unifying 
all interactions and such unification is our original intention since
 {\it our job in physics is to see things simply, to understand a great many
complicated phenomena in a unified way \cite{weinberg}}.

Of course, {\it physics thrives on crisis \cite{weinberg-2}}. 
Since particle physics is a discipline relying on experiments, the experimental crises 
or deviations from the SM play a crucial role in searching for new physics.
Currently, we are facing two experimental crises: one is the cold dark matter, the other is 
the muon $g-2$ anomaly. While the former is a quite robust evidence of new physics, the latter 
should be taken with a grain of salt, e.g., {\it 10\% chance it is new physics (much more 
plausible than other anomalies) \cite{nima-talk}}. In addition, the dark matter direct detection experiment has not
found any WIMP (Weakly Interactive Massive Particle) dark matter particle.  The LHC has not found any particles predicted by new physics
(albeit found some plausible anomalies in B-decays).     
Confronted with these experimental results, what is the status of low energy SUSY, healthy or
needs to be hospitalized ? 

In this note, we will briefly review the status of SUSY in light of these experiments, namely 
the LHC searches, the dark matter, and muon $g-2$ measurements. The SUSY models investigated in the literature are the MSSM, mSUGRA or CMSSM, GMSB, and AMSB, as well as respective extensions. 
We will conclude that the low energy SUSY can survive all these experimental 
constraints, albeit the fancy models like mSUGRA, GMSB and AMSB need to be extended 
to accommodate the 125 GeV Higgs boson and the muon $g-2$ anomaly.
       
\section{A brief description of SUSY}
{\it Supersymmetry is an extension of special relativity to include fermionic symmetries \cite{witten-bj}}.
The anticommutative relation of the supercharges $Q_\alpha$ is given by \cite{witten-bj}
\begin{equation}
Q_\alpha Q_\beta+Q_\beta Q_\alpha=\Gamma^\mu_{\alpha\beta} P_\mu ,
\end{equation}
with $\Gamma^\mu$ being the Dirac matrix and $P_\mu$ the four-momentum. 
So, unlike any other internal symmetries which are independent of spacetime symmetry
(no-go theorem of Coleman and Mandula \cite{no-go}: The most general Lie algebra of symmetry operators 
that commute with S-matrix consists of the generators of the Poincare group and ordinary 
internal symmetry generators. The latter act on one-particle states with matrices that are 
diagonal and independent of both momentum and spin), 
supersymmetry is entangled with spacetime symmetry and is an extension of special relativity
(Golfand and Likhtman \cite{golfand} found that the S-matrix can have Poincare symmetry extended by 
SUSY algebra,
while Haag et al. \cite{haag} further proved that SUSY algebra is the only graded Lie algebra of symmetries 
of the S-matrix consistent with relativistic QFT, which extends the Poincare group by anti-commutators).
In order words, the spacetime symmetry of QFT is completed by SUSY: 
\begin{eqnarray}
&& \left [ P_\mu, P_\nu \right]=0 ,\\
&& \left [ M_{\mu\nu}, P_\lambda \right] 
  = i\left (  P_\mu ~g_{\nu\lambda}-  P_\nu ~ g_{\mu\lambda} \right ) ,\\
&& \left [ M_{\mu\nu}, M_{\rho\lambda} \right] =-i  
      \left ( M_{\mu\rho} ~g_{\nu\lambda} + M_{\nu\lambda} ~g_{\mu\rho}
              - M_{\mu\lambda} ~g_{\nu\rho} - M_{\nu\rho} ~g_{\mu\lambda}\right ) ,\\
&& \left [ P_\mu,  Q_a \right]=0 ,\\
&& \left [ M_{\mu\nu}, Q_a \right]=-\left(\sigma_{\mu\nu}^4\right)_{ab} Q_b ,\\
&& \left \{ Q_a, Q_b \right\}= -2 \left( \gamma^\mu C\right)_{ab} P_\mu ,\\
&& \left \{ \bar Q_a, \bar Q_b \right\}= 2 \left(C^{-1} \gamma^\mu \right)_{ab} P_\mu , \\  
&& \left \{ Q_a, \bar Q_b \right\}= 2 \left(\gamma^\mu \right )_{ab} P_\mu , 
\end{eqnarray}  
where $P_\mu$ and  $M_{\mu\nu}$ are the generators of the Poincare group.    
At about the same time as Golfand and Likhtman,  a string was introduced that could be a fermion
as well as a boson led to the concept of supersymmetry \cite{ramond,neveu,sakita}. 
When supersymmetry was applied to particle physics (pioneered by Wess and Zumino \cite{wess}), 
it demonstrated remarkable virtues such as \cite{witten}:
makes a ``small'' Higgs mass natural, survives electroweak tests, makes heavy top mass as needed. 
Further, when supersymmetry is localized, we obtain supergravity \cite{supergravity}. 

Since any field representation of SUSY algebra involves fields with
different spins and the same mass, SUSY predicts a superpartner for each particle in the SM. 
If SUSY is unbroken, we will find light sparticles that were ruled out by experiments. So SUSY must be 
spontaneously broken in some hidden sector where the breaking effects can be mediated to 
the observable sector via some interactions. Note that the supertrace sum 
rule \cite{Martin,baer-book}
\begin{eqnarray}
Str M^2=\sum_{particles} (-1)^{2j} (2j+1) M_j^2 =0 ~, 
\end{eqnarray} 
with $j$ being the spin, which holds for any renormalizable SUSY Lagrangian,  will lead to the mass relation between electron and selectrons 
\begin{eqnarray}
m_{\tilde e_1}^2+m_{\tilde e_2}^2  =2 m_e^2~,
\end{eqnarray}  
indicating the existence of a selectron lighter than the electron.  Hence SUSY breaking cannot be merely like the spontaneous breaking of electroweak symmetry in the Higgs sector, 
which directly couples to the fermion and gauge fields to transfer the breaking effects.
Suppose the SUSY breaking in the hidden sector is characterized by the F-term VEV $\langle F\rangle$ and the mediation scale is $M_{med}$, 
the superpartners in the visible sector obtain a soft mass from SUSY breaking :
\begin{eqnarray}
M_{soft} \sim  \frac{\langle F\rangle}{M_{med}} ~,
\end{eqnarray}  
up to some numerical factor such as the loop factor for gauge mediation.  $M_{med}$  can be identified to be the Planck scale $M_{Pl}$ for gravity mediation and  the messenger scale $M_{mess}$ for gauge mediation, with $M_{mess}$  possibly being much smaller than the Planck scale $M_{mess}\ll M_{Pl}$. 
{\it Superpartners of the standard model particles  get masses both from  electroweak symmetry breaking and SUSY breaking. Therefore it is
natural for them to be a bit above the $Z$, which gets mass only from electroweak symmetry breaking \cite{witten}}.
    
Although SUSY predicts numerous sparticles, only the lightest sparticle (LSP)  
is stable assuming R-parity (without R-parity no sparticles are stable), which is usually assumed 
to be the lightest neutralino. This lightest neutralino is a good candidate for the WIMP cold 
dark matter, and at the LHC it appears as missing energy at the end of the decay chain of each 
produced sparticle.  

\begin{figure}[htb]
\begin{center}
\includegraphics[width=7cm, height=7cm]{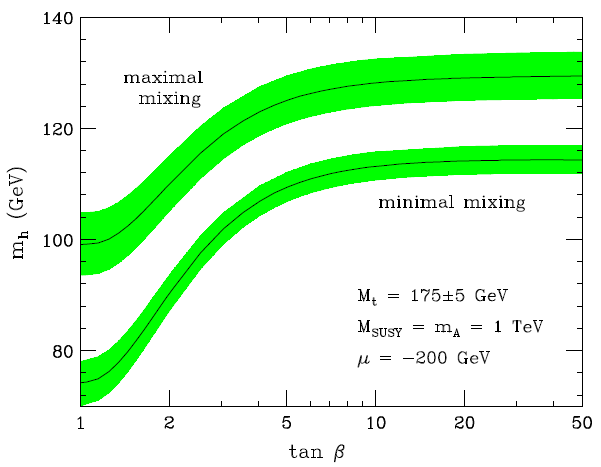} \includegraphics[width=8cm, height=7cm]{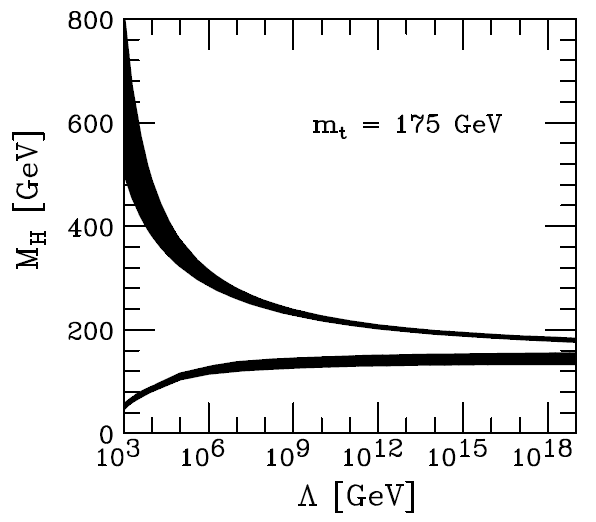}
\end{center}
\vspace{-.5cm}
\caption{The SM-like Higgs boson mass range in the MSSM taken from \cite{HiggsWorkingGroup:2000zik}, 
 compared with the SM Higgs boson mass range taken from \cite{Riesselmann:1997kg} which considered  
the requirements of non-triviality \cite{Hambye:1996wb}
and vacuum stability \cite{Altarelli:1994rb,Casas:1994qy}. }
\label{higgs-mass}
\end{figure}

Another remarkable feature of SUSY is that to give masses for both up and down type quarks, two Higgs
doublets with opposite hypercharges are needed since SUSY forbids the appearance of the complex 
conjugate of the Higgs field in the superpotential. So SUSY predicts five Higgs bosons: two CP-even
ones, one CP-odd one, and a pair of charged ones.  The Higgs quartic coupling 
$\lambda$ arises from D-terms \cite{Gunion:1984yn}
\begin{eqnarray}
V_D=\frac{1}{2}\left[ D^a D^a +(D')^2 \right ] 
   =\frac{1}{8}(g^2+g'^2) \left(H_d^\dagger H_d-H_u^\dagger H_u\right)^2
    +  \frac{1}{2}g^2 \left(H_d^\dagger H_u \right)^2 , 
\end{eqnarray}
 which is gauge coupling $\lambda\sim g^2$ 
instead of a free parameter as in the SM.  Thus in the MSSM, the mass 
of the lightest CP-even Higgs boson must be light $m_h^2 \sim \lambda v^2 \sim g^2 v^2$. 
In detail, its mass is upper bounded by $Z$-boson mass at tree level and by about 135 GeV 
at loop level: 
\begin{eqnarray}
&& m_h < m_Z|\cos2\beta| <  m_Z ~~~~~~~~ ({\rm at~ tree ~level}) , \\
&& m_h \le \sqrt{m_Z^2 +\epsilon} \le 135~ {\rm GeV}  ~~~~ ({\rm at~ loop ~level}),
\label{135}
\end{eqnarray}
 where  $\epsilon$ is the one-loop effects from top quarks and top squarks given by  \cite{Carena:2011aa}
\begin{eqnarray}
\epsilon =  \frac{3m_t^4}{4\pi^2 v^2}  
  \left[ \log\frac{M_S^2}{m_t^2}+\frac{X_t^2}{M_S^2}\left(1-\frac{X_t^2}{12 M_S^2}\right) \right] ,
\end{eqnarray}  
with  $v=174 {\rm ~GeV}$,  $X_t \equiv A_t - \mu \cot \beta$ ($A_t$ is the trilinear Higgs-stop coupling 
and $\mu$ is the Higgsino mass parameter), and  
$M_S = \sqrt{m_{\tilde{t}_1}m_{\tilde{t}_2}}$. 
As shown in Fig.\ref{higgs-mass}, the SM-like Higgs boson mass in the MSSM is more restricted than 
the  SM Higgs boson mass. Since SM Higgs boson mass is bounded by about 800 GeV if its cut-off scale is 1 TeV.
 
\section{SUSY confronted with the LHC searches}

\begin{figure}[htb]
\begin{center}
\includegraphics[width=7.2cm]{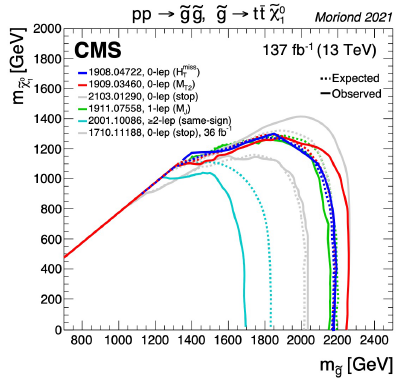}  
\includegraphics[width=7cm]{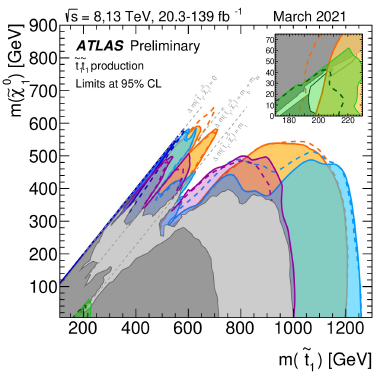}
\end{center}
\vspace{-.5cm}
\caption{The current LHC bounds on gluino and stop masses,  taken from \cite{susy-moriond-2021}.}
\label{susy-lhc}
\end{figure}

At the LHC the typical signature of SUSY with R-parity is multi-jets or/and multi-leptons plus missing energy
since the sparticles must be produced in pairs and each produced sparticle has a cascade decay with the final
states containing an odd number of LSP (the stable neutral LSPs just escape the detector) plus 
some jets or/and leptons~\cite{Han:2016xet,Kobakhidze:2015scd,Han:2013kga,Cao:2012rz}.  
So far the CMS and ATLAS collaborations at the LHC have intensively searched for sparticle productions 
and have failed to find any evidence. For simplified models with significant mass splittings between 
the LSP and other sparticles, the searches have pushed gluino and first generation squarks above about 2 TeV 
\cite{gluino-cms,gluino-atlas}, pushed top squarks above about 1 TeV \cite{stop-atlas,stop-cms}, 
as shown in Fig.\ref{susy-lhc},  
while for uncolored sparticles (electroweakinos and sleptons) the bounds are much weaker, 
only a few hundred GeV due to their small production cross-sections 
\cite{ATLAS:2020pgy,ATLAS:2019lff,ATLAS:2019wgx}.  
As we will discuss later, these relatively light  uncolored sparticles are just needed to make sizable contributions to the muon $g-2$.    
 
\begin{figure}[htb]
\begin{center}
\includegraphics[width=14cm]{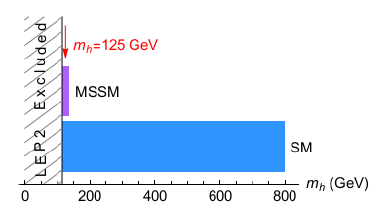}
\end{center}
\vspace{-.5cm}
\caption{The Higgs mass ranges predicted by the MSSM and SM compared with the discovered 125 GeV 
Higgs mass, taken from \cite{baer}. }
\label{mssm-125}
\end{figure}
\begin{figure}[htb]
\begin{center}
\includegraphics[width=10cm,height=9cm]{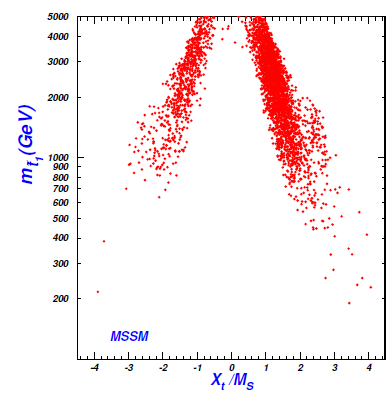}
\end{center}
\vspace{-.5cm}
\caption{The scatter plots of the parameter space in the MSSM 
giving $125\pm 2$ GeV Higgs mass and satisfying other constraints, 
taken from \cite{Cao:2012fz}. }
\label{mssm-stop}
\end{figure}

On the other hand, the LHC discovered a 125 GeV Higgs boson,
which is within the range $< 135$ GeV predicted by SUSY as shown in Fig.\ref{mssm-125} \cite{baer}, 
and requires sizable loop effects of top squarks heavier than about TeV for 
 $X_t \equiv A_t - \mu \cot \beta$ not much larger than $M_S \equiv \sqrt{m_{\tilde t_1} m_{\tilde t_2}}$, 
as shown  in  Fig.\ref{mssm-stop} \cite{Cao:2012fz}. 
If we believe the stops to be 
the lightest colored sparticles from the view of renormalization-group equation (RGE) runnings, the 125 GeV Higgs mass 
then requires all colored sparticles above TeV.  So the absence of any colored sparticles 
and the discovery of a 125 GeV Higgs boson is consistent in the framework of SUSY, both pushing 
the colored sparticles above TeV. Whereas, since the uncolored sparticles are subject to extremely weak 
constraints from the LHC searches, their allowed low masses below TeV are just required by the muon $g-2$ discrepancy, as discussed later.

Note that the LHC has not yet discovered any particles predicted by new physics, not just the sparticles 
predicted by SUSY. Among the new physics models, SUSY remains the most compelling one. 
Another point we should note is that the LHC bounds on the sparticle masses are usually valid for 
simplified SUSY models with significant mass splittings between the LSP and other sparticles.
If the LSP mass is very close to the relevant sparticles like the higgsinos \cite{higgsino-lsp} 
or stops \cite{stop-lsp-1,stop-lsp-2}, 
the LHC bounds on the masses of these sparticles will become much weaker.    
For example, in the natural SUSY scenario 
\cite{naturalness-1,naturalness-2,naturalness-3,naturalness-4,Baer:2012up,Tata:2020afe}
the higgsinos are light 100-300 GeV while the gauginos are heavy, and thus
the LSP $\chi^0_1$ is higgsino-like and nearly degenerate with the  higgsino-like neutralino  $\chi^0_2$ 
and the higgsino-like chargino  $\chi^{\pm}_1$. In this case, the productions of these sparticles  
$\chi^0_{1,2}$ and  $\chi^{\pm}_1$ at the LHC give missing energy and can only be searched by 
requiring a hard jet radiated from initial partons, i.e., the signal of monojet plus missing 
energy \cite{higgsino-lsp}. Moreover, global likelihood analysis of the electroweakino sector shows that no range of neutralino
or chargino masses can be robustly excluded by current LHC searches\cite{GAMBIT:2018gjo}. 

\section{SUSY confronted with the dark matter}
The cold dark matter can be explained in SUSY by the neutral and stable LSP which can be a weakly interacting massive particle (WIMP)-the lightest neutralino or a superweakly interacting massive particles (superWIMP) such as the pseudo-goldstino or gravitino: 
\begin{itemize}  
\item[(i)] 
For the lightest neutralino case, it can be bino-like, higgsino-like, or wino-like, depending on its 
dominant components. 
The bino-like LSP can easily satisfy the direct detection limits since its scattering with the nucleon
is very weak and also can give the correct thermal relic density through the coannihilation with 
other sparticles like winos \cite{bino-wino} or sfermions. For the  bino-like dark matter in the MSSM,
the parameter space satisfying the relic density at $2\sigma$ level (and also explaining the muon $g-2$
 at $2\sigma$ level) is displayed in Fig.\ref{mssm-dm} which is taken from \cite{gm2-mssm}
(see also  \cite{gm2-mssm-2,gm2-mssm-3,gm2-mssm-4,gm2-mssm-5,gm2-mssm-6,gm2-mssm-7}). It has to be heavier than 100 GeV, owing to the LHC direct searches of electroweakino and the DM direct searches\cite{Pozzo:2018anw}.
For the bino-like dark matter in the CMSSM/mSUGRA,
the $2\sigma$ region from a global fit considering various measurements including the dark matter relic 
density is shown in Fig.\ref{cmssm-dm} taken from \cite{fit-cmssm} (also see \cite{fit-cmssm-2} for the most updated global fit of CMSSM).   
  
The higgsino-like or wino-like LSP has relatively strong interactions and freeze out rather late 
from the thermal bath in the early universe, leading to an under-abundance of dark matter if 
they are light below TeV. 
To give the sufficient aboundance for the dark matter, the thermal freeze-out higgsino-like (wino-like) 
LSP must be around $1.0\pm 0.1$ TeV \cite{marco,hisano} ($2.9\pm 0.1$ TeV \cite{hisano,wino-lsp2,wino-lsp3}), 
which is so far allowed by the direct or 
indirect detections and will be sensitively probed by the indirect detections like the upcoming 
Cherenkov Telescope Array \cite{Rinchiuso:2020skh}.  

The light higgsino-like LSP with an under-abundance
can also satisfy the direct detection limits. For example, in natural SUSY the higgsinos are quite light,
around 100-300 GeV, and thus the thermal relic density of the higgsino-like LSP is far below the dark
matter abundance. If we want to enhance its thermal relic density to the required abundance by 
mixing bino with higgsino, it may lead to a too large scattering cross-section with the nucleon 
\cite{mulat-bino-higgsino}. 
\begin{figure}[htb]
\begin{center}
\includegraphics[width=10cm,height=8cm]{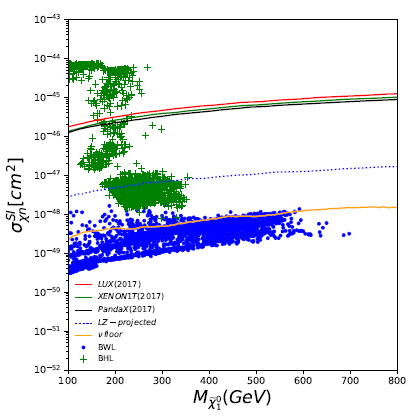}
\end{center}
\vspace{-.5cm}
\caption{The scatter plots of the bino-like dark matter parameter space in the MSSM satisfying 
the relic density at $2\sigma$ level (and also explaining the muon $g-2$ at $2\sigma$ level),
displayed in the plane of the direct detection limits on the spin-independent scattering 
cross section with the nucleon, taken from \cite{gm2-mssm}.}
\label{mssm-dm}
\end{figure}   
\begin{figure}[htb]
\begin{center}
\includegraphics[width=10cm,height=7.8cm]{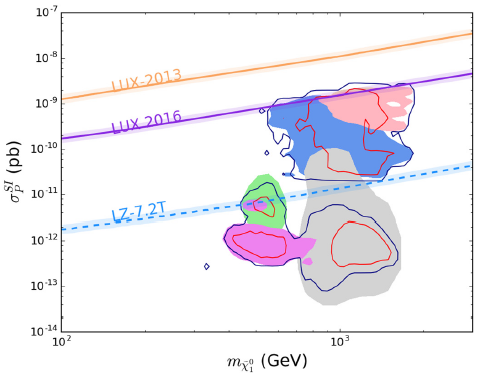}
\end{center}
\vspace{-.5cm}
\caption{The $2\sigma$ region of the bino-like dark matter parameter space in the CMSSM/mSUGRA 
satisfying the relic density and other observables at $2\sigma$ level (not considering the muon $g-2$),
displayed in the plane of the direct detection limits on the spin-independent scattering cross section 
with the nucleon, colored by DM annihilation mechanisms, taken from \cite{fit-cmssm}.}
\label{cmssm-dm}
\end{figure}

\item[(ii)] For the superWIMP pseudo-goldstino or gravitino, 
it can be produced from the late decay ~\cite{Feng:2004mt, Wang:2004ib, Gu:2020ozv} 
of the freeze-out lightest neutralino and thus can provide 
the correct dark matter relic density while easily satisfying the direct detection limits.  
The pseudo-goldstinos are predicted in multi-sector SUSY breaking with different scales, in which 
amount of goldstinos is generated, with one linear combination of these goldstinos being massless 
and eaten by the gravitino while the orthogonal combinations acquire masses and become the physical 
states (depending on the messenger masses in the context of gauge mediation, the pseudo-goldstino mass 
can be from 0.1 GeV to hundred GeV)~\cite{Argurio:2011hs,Dai:2021eah}. 
Note that if the superWIMP is thermally produced in the reheating 
 period after inflation, its relic density could be overabundant and 
hence set an upper bound on the reheating temperature \cite{Moroi:1993mb}. 
Such superWIMP dark matter is usually not accessible in direct or indirect
detections, but may cause exotic phenomenology at the colliders and may help to alleviate some cosmic problem, such as the Hubble tension~\cite{Wu:gravitino} or Xenon1T excess~\cite{Fei:2007.09981}. The  produced lightest ordinary supersymmetric particle (LOSP) at the colliders, which can possibly be charged,
could decay to the superWIMP plus visible particles (photon, Higgs boson,  Z-boson etc) inside or outside the 
detectors \cite{Liu:2013sx,Hikasa:2014yra,Liu:2014lda,Franzosi:2021zwp,Chen:2021omv}, depending on the interaction strength of the superWIMP particle.         
\end{itemize}  
Therefore, SUSY (both the phenomenological MSSM and the constrained frameworks like the mSUGRA) 
can explain the cold dark matter relic density and satisfy the tightened direct detection 
limits on the WIMP.  
 
\section{SUSY confronted with the muon $g-2$} 
The muon $g-2$ from the Fermilab  \cite{FNAL:gmuon} measurement combined with the 
BNL result \cite{BNL:gmuon} shows a $4.2\sigma$ deviation from the SM prediction \cite{sm-prediction-1,sm-prediction-2} 
(however, if the lattice simulation result of the hadronic contribution from the BMW group is taken,  the deviation can be reduced to  $1.5\sigma$ \cite{gm2-lattice}), which can be readily explained in the phenomenological minimal supersymmetric model (MSSM). The explanation requires relatively light uncolored sparticles (sleptons and electroweakinos).
While for the models with boundary conditions at some UV scales for the soft parameters, such as the mSUGRA, GMSB, they need to be extended to account for the muon $g-2$ and the 125 GeV Higgs boson mass. 

In SUSY, the smuons, muon sneutrino, and electroweakinos can contribute to muon $g-2$ at the one-loop level. 
Since the SUSY contribution to  muon $g-2$ \cite{moroi,stockinger} is enhanced by a large $\tan\beta$ 
and suppressed by heavy sparticle masses involved in the loops,
to generate the required contribution to explaining the muon $g-2$ deviation, a low
SUSY mass and a large $\tan\beta$ are favored.  
\begin{itemize}

\item[(i)] In the low energy effective MSSM, the masses of bino, winos, higgsinos, smuons and sneutrino 
are all independent parameters. As shown in Fig.\ref{mssm-gm2} taken from \cite{gm2-gutsusy} 
(see also \cite{gm2-gutsusy-2,gm2-gutsusy-3}), 
under other constraints including the dark matter, 
the vacuum stability and the LHC search for sleptons, there still remains a part of 
the MSSM parameter space which can explain the muon $g-2$ at $2\sigma$ level 
\cite{gm2-gutsusy,gm2-gutsusy-2,gm2-gutsusy-3,gm2-mssm}.

\begin{figure}[htb]
\begin{center}
\includegraphics[width=10cm]{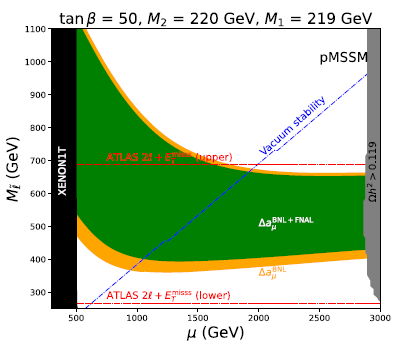}
\end{center}
\vspace{-.5cm}
\caption{ The MSSM parameter space for the explanation of the muon $g-2$ at $2\sigma$ 
level,taken from  \cite{gm2-gutsusy}. The upper-left green region above the 
dashed line is allowed by all constraints and can explain the muon $g-2$ at $2\sigma$ 
level.}
\label{mssm-gm2}
\end{figure}
\begin{figure}[htb]
\begin{center}
\includegraphics[width=10cm]{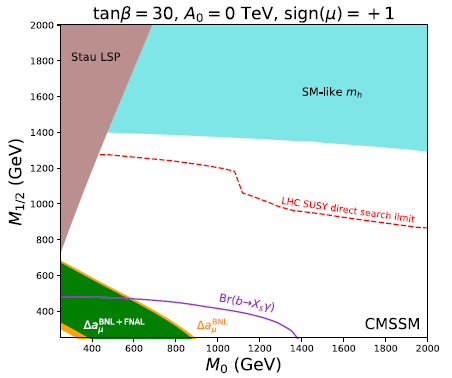}
\end{center}
\vspace{-.5cm}
\caption{The tension between the muon $g-2$ explanation and the 125 GeV SM-like Higgs mass 
as well as the LHC searches for the CMSSM/mSUGRA,  taken from  \cite{gm2-gutsusy}. }
\label{tension-cmssm}
\end{figure}
\begin{figure}[htb]
\begin{center}
\includegraphics[width=12cm]{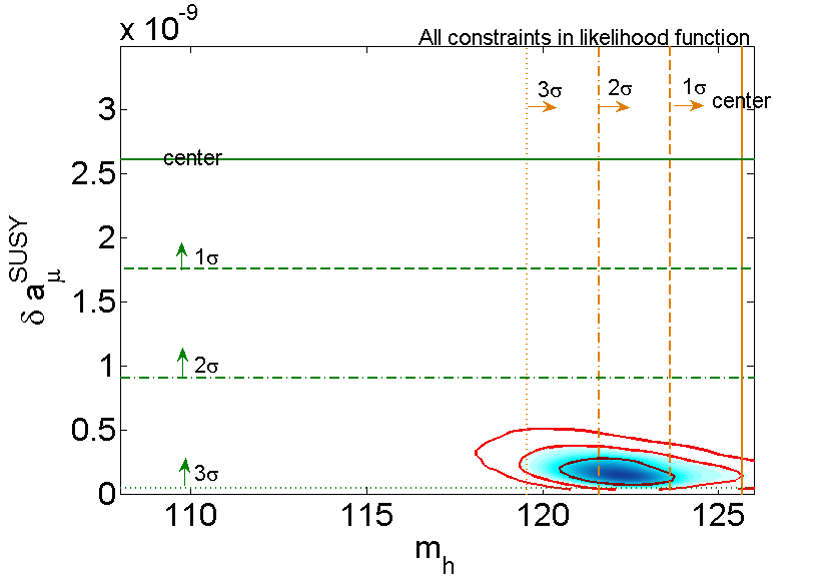}
\end{center}
\vspace{-.5cm}
\caption{The tension between the muon $g-2$ and the 125 GeV SM-like Higgs mass for the CMSSM/mSUGRA, plotted from a global fit in \cite{fit-cmssm}. 
The regions encircled by the curves correspond to the $1\sigma$, $2\sigma$  and $3\sigma$ levels,
respectively.}
\label{tension-cmssm-fit}
\end{figure}

\item[(ii)]  In the constrained models with certain UV boundary conditions for the soft breaking parameters,
such as mSUGRA/CMSSM, GMSB,  the soft masses at the weak scale are correlated. 
To give a 125 GeV SM-like Higgs mass, the top squarks must be heavy and the correlated slepton masses cannot be as light
as required by the explanation of the muon $g-2$ anomaly. 
Fig.\ref{tension-cmssm} and Fig.\ref{tension-cmssm-fit} 
show the tension between the muon $g-2$ explanation and the 125 GeV SM-like Higgs mass 
as well as the LHC searches for the CMSSM/mSUGRA \cite{gm2-gutsusy} with inputs at the GUT scale.  
To solve such a tension, these models need to be extended, e.g.,  couple the messengers with 
the Higgs doublets to raise the tree-level SM-like Higgs mass 
\cite{GMSB-yukawa-higgs-1,GMSB-yukawa-higgs-2} in gauge mediation or
make the colored sparticles much heavier than 
uncolored sparticles at the weak scale 
\cite{gm2-SUGRA-ext-1,gm2-SUGRA-ext-2,gm2-SUGRA-ext-3,gm2-SUGRA-ext-4,gm2-SUGRA-ext-5,gm2-SUGRA-ext-6}. 
In superGUT~\cite{super:GUT} or subGUT~\cite{sub:GUT} scenarios for SUGRA, which adopts the UV input upon or below the GUT scale, there are still survived parameter spaces that can account for both the muon g-2 anomaly and other constraints, such as the dark matter and collider searches.  In the deflected AMSB, which is an elegant extension of AMSB, the muon g-2 anomaly can easily be explained~\cite{Fei:1508.01299, Fei:1512.06715,Fei:1703.10894,Fei:1710.06105,Fei:1903.05669}.   The mixed mediation scenarios, such as the mirage mediation, can also successfully account for the muon g-2 anomaly~\cite{Fei:1804.07335, Fei:1808.08529}.   
\item[(ii)] The non-minimal frameworks of SUSY with more free parameters than the MSSM 
can easily accommodate the 125 GeV 
SM-like Higgs mass and the muon $g-2$ at $2\sigma$ level, e.g., the next-to-minimal 
SUSY model (NMSSM) extends the MSSM by a singlet Higgs superfield and can satisfy 
all current constraints \cite{gm2-nmssm-1,gm2-nmssm-2,gm2-nmssm-3,Wang:2021lwi}.      
\end{itemize}

We should also note the electron $g-2$ which shows a slight deviation according to the 
Berkeley measurement \cite{berkeley}. A joint explanation of 
such an electron $g-2$ and the muon $g-2$ is feasible in the MSSM \cite{joint-1,joint-2,joint-3,joint-4,joint-5,Ali:2021kxa}. 
If we further consider the measurement of electron EDM, its correlation 
with the muon $g-2$ in the MSSM can set stringent constraints on the CP-phases 
of the soft parameters \cite{cp-phase-1,cp-phase-2}.
  
In this review, we did not discuss the plausible B-decay 
anomalies, which are hard to be explained in SUSY with minimal flavor violation and the explanation needs non-minimal flavor violation \cite{b-decay-susy}.

\section{What is the problem of SUSY}
{\it The most obvious drawback is simply that SUSY has not been found yet, though we have 
been hoping for a long time. It is disappointing that we have not found SUSY yet, but for the 
most part, it is perhaps not too surprising \cite{witten}}. 
Although the undiscovery of sparticles at the LHC is not too surprising because sparticles
obtain masses from both electroweak symmetry breaking and SUSY breaking and should be significantly 
heavier than their SM partners,      
the top squarks which have been pushed above TeV by the direct LHC searches and by the 125 GeV 
Higgs boson mass will unpleasantly enlarge the logarithmic divergence in the Higgs boson mass
and lead to a little fine-tuning \cite{Papucci:2011wy, Baer:2016bwh}: 
\begin{eqnarray}
\Delta_{\rm HS} \equiv\frac{\delta m_h^2}{m_h^2}=\frac{3y_t^2}{4\pi^2m_h^2}(m_{Q_3}^2+m_{U_3}^2
+A_t^2)\log\frac{\Lambda}{M_{\rm SUSY}} ,
\label{naturalness}
\end{eqnarray}
with $M_{\rm SUSY}=\sqrt{m_{\tilde{t}_1}m_{\tilde{t}_2}}$, $\Lambda$ being the cut-off scale,
$Q_3=(\tilde{t}_L,\tilde{b}_L)$ and $U_3=\tilde{t}_R$. 
\begin{figure}[htb]
\begin{center}
\includegraphics[width=13cm]{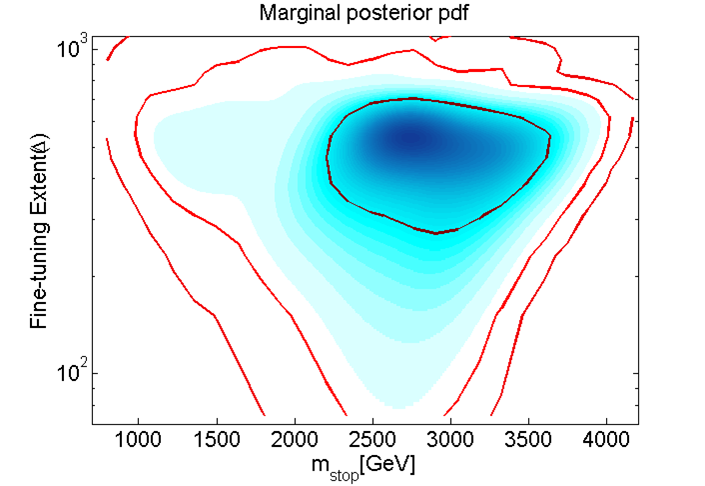}
\end{center}
\vspace{-.5cm}
\caption{The marginal posterior probability distribution function of the CMSSM showing 
the tuning extent versus the stop mass from a Bayesian analysis plotted from the analysis 
in \cite{fit-cmssm} where the muon $g-2$ data is not considered. }
\label{tuning-cmssm}
\end{figure}

The more traditional measure of tuning extent is defined by \cite{Barbieri:1987fn,Ellis:1986yg} 
\begin{equation} 
\Delta_{\rm BG}
\equiv max_i\left|\frac{p_i}{m_Z^2}\frac{\partial m_Z^2}{\partial
  p_i}\right| ,
\label{eq:DBG}
\end{equation}
with $p_i$ being the independent SUSY parameters at the cut-off scale. 
In terms of this measure, in the CMSSM the fine-tuning extent is about per mille, as shown in 
Fig.\ref{tuning-cmssm}  \cite{fit-cmssm}. The fine-tuning  can be lower \cite{baer} if we use another measure 
$\Delta_{\rm EW}$ \cite{Baer:2012up,Baer:2013gva}:
\begin{equation}  
\Delta_{\rm EW} \equiv max_i \left|C_i\right|/(m_Z^2/2) ,
\end{equation}
with \cite{Baer:2012cf} 
\beqa
C_{H_u}&=&\left|-\f{m^2_{H_u}\tan^2\beta}{\tan^2\beta-1}\right|~,~
C_{H_d}=\left|\f{m^2_{H_d}}{\tan^2\beta-1}\right|~,~
C_{\mu}=\left|-\mu_{eff}^2\right|~,~\nn\\
C_{\Sigma_{u}^u(\tl{t}_{1,2})}&=&
\f{\tan^2\beta}{\tan^2\beta-1}\left|\f{3}{16\pi^2}F(m_{\tl{t}_{1,2}}^2)\left[y_t^2-g_Z^2\pm \f{y_t^2 A_t^2-8g_Z^2(\f{1}{4}-\f{2}{3}x_w)\Delta_t}{m_{\tl{t}_{2}}^2-m_{\tl{t}_{1}}^2}\right]\right|~,\nn\\
C_{\Sigma_{d}^d(\tl{t}_{1,2})}&=&\f{1}{\tan^2\beta-1}\left|\f{3}{16\pi^2}F(m_{\tl{t}_{1,2}}^2)\left[g_Z^2\pm \f{y_t^2 \mu_{eff}^2+8g_Z^2(\f{1}{4}-\f{2}{3}x_w)\Delta_t}{m_{\tl{t}_{2}}^2-m_{\tl{t}_{1}}^2}\right]\right|~,
\eeqa
where $x_w=\sin^2\theta_W$ and
\beqa
\Delta_t&=&\f{(m_{\tl{t}_{L}}^2-m_{\tl{t}_{R}}^2)}{2}+M_Z^2\cos2\beta(\f{1}{4}-\f{2}{3}x_w)~,\nn\\
F(m^2)&=& m^2 \(\log\frac{m^2}{m_{\tilde{t}_{1}}m_{\tilde{t}_{2}}}-1\)~.\eeqa
Anyway, we admit that SUSY has a little fine-tuning.  
Note that a solution has been developed to solve such a little fine-tuning problem \cite{Cohen:2020ohi}, in which a framework of supersoft top-squarks is proposed
to soften the logarithms to screen the UV-sensitive logs.
    
Another unpleasant point caused by top-squarks above TeV is that the MSSM can no longer realize the 
first order electroweak phase transition. However, the NMSSM can do this job 
(see \cite{Huang:2014ifa} and refs therein).   

\begin{figure}[htb]
\begin{center}
\includegraphics[width=15cm]{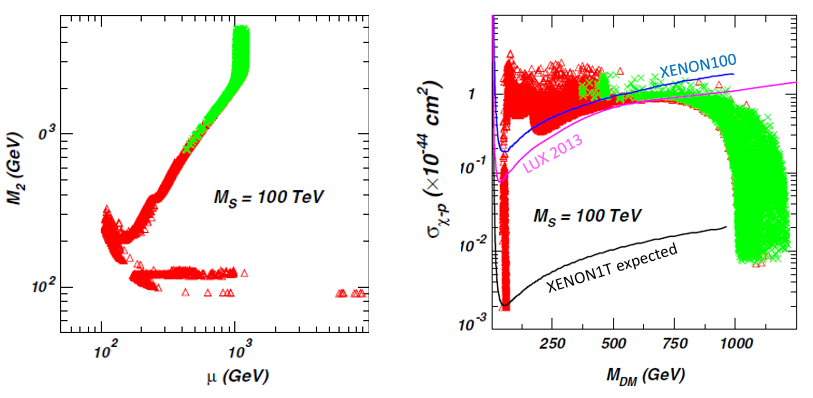}
\end{center}
\vspace{-.5cm}
\caption{The parameter space of split-SUSY satisfying dark matter relic density ($2\sigma$ range of Planck data) and the $125\pm 2$ GeV Higgs mass,
taken from \cite{Wang:2013rba}. 
The green (red) samples can (cannot) achieve the gauge coupling unification at GUT scale. 
}
\label{split-susy}
\end{figure}

Note that in this review we focused on low energy SUSY which helps to solve the naturalness problem in particle physics. It is clear that even with low energy SUSY our particle theory still has a little tuning, which means we cannot have perfect naturalness. If we give up naturalness and believe simplicity, we may have the fascinating split-SUSY \cite{split-1,split-2,split-3,Wang:2015mea}, which is emphasized in the recent Witten Reflects \cite{witten-reflect}. 
Split-SUSY gives up naturalness and retains the original motivation to explain the dark matter and realize successful gauge coupling unification.  
 Actually, as shown in Fig.\ref{split-susy}, the dark matter relic density and direct detection limits as well as the gauge coupling unification impose stringent constraints on the parameter space of split-SUSY, i.e., only the higgsino-like dark matter around 1.2 TeV can survive the XENON1T limits assuming the universal GUT input for gaugino masses (which leads to $M_1:M_2:M_3\sim ~1:~2:~6$ at the weak scale).   
Since in split-SUSY the gauginos and higgsinos are not so heavy,
they may be accessible at future colliders \cite{Wang:2013rba,Wang:2005kf}.
{\it This is that the electroweak scale is not natural
in the customary sense, but additional particles and forces that would help us understand what is going on exist at an energy not too much above LHC energies \cite{witten-reflect}}.  

\section{The continual search of SUSY at HL-LHC}
Despite the fact that no sparticles have been discovered so far, SUSY will be actively probed at the High-Luminosity Large Hadron Collider (HL-LHC) as the leading contender for new physics beyond the SM.

\begin{figure}[htb]
\begin{center}
\includegraphics[width=10cm,height=12cm]{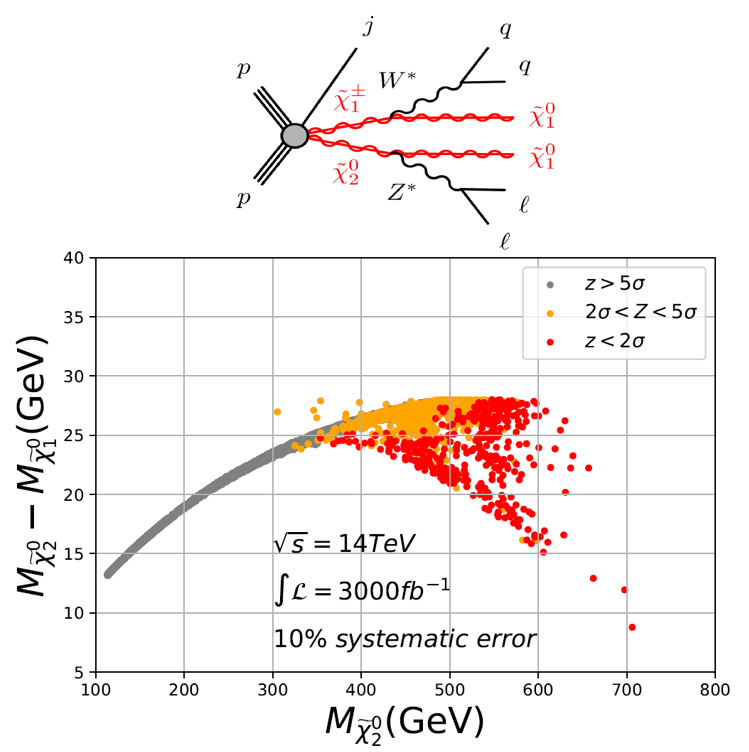}
\end{center}
\vspace{-.5cm}
\caption{ The HL-LHC coverage of the MSSM parameter space for the explanation of the 
muon $g-2$ at $2\sigma$ level, taken from \cite{gm2-mssm}. }
\label{hl-lhc-mssm}
\end{figure}
\begin{figure}[htb]
\begin{center}
\includegraphics[width=13cm]{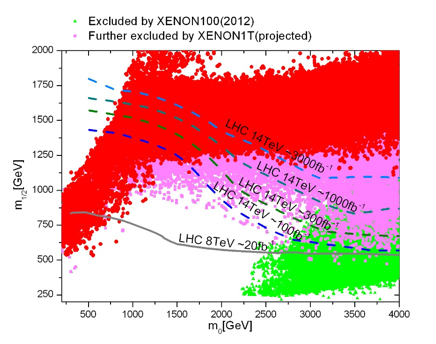}
\end{center}
\vspace{-.5cm}
\caption{The $2\sigma$ parameter space of the CMSSM from a global fit, showing 
the coverages of the current and future LHC runs, plotted from the analysis 
in \cite{fit-cmssm}.}
\label{hl-lhc-cmssm}
\end{figure}
As shown above, the MSSM can allow for an explanation for the muon $g-2$ at $2\sigma$ level while
satisfying all other constraints.  
The light electroweakinos and sleptons required for the muon $g-2$ explanation may be accessible at the HL-LHC. For example, in the scenario where bino-like dark matter 
annihilates with winos to give the correct dark matter abundance, the light winos required by
the muon $g-2$ may be fairly produced and detected through the signal of soft leptons plus missing energy  
at the LHC, as shown in Fig.\ref{hl-lhc-mssm} \cite{gm2-mssm}. 
The $2\sigma$ parameter space is shown in this figure was allowed by   
the muon $g-2$, the dark matter relic density, and direct detection limits as well as 
the LHC Run-2 data  \cite{gm2-mssm}. 
We see that the HL-LHC can cover a sizable part of the MSSM parameter space for the explanation 
of muon $g-2$ at $2\sigma$ level. In addition, the precision measurement of the Higgs properties 
at a lepton collider as a Higgs factory could play a complementary role via detecting the 
residual effects of SUSY in Higgs couplings  \cite{gm2-mssm, Li:2020glc}. 

Since the muon $g-2$ may not be a real hint of new physics, we consider the CMSSM without 
explaining the muon $g-2$. As shown in Fig.\ref{hl-lhc-cmssm}, the HL-LHC will continue
to narrow down the CMSSM parameter space. The $2\sigma$ parameter space  shown in this figure 
was obtained from a likelihood analysis, taking into account various data including the electroweak 
precision observables, the $B$-physics measurements, the LHC Run-1 and Run-2 data of 
SUSY direct searches, the Planck observation of the dark matter relic density and the 
combined LUX Run-3 and Run-4 detection limits \cite{fit-cmssm}. 

\section{Conclusion} 
So we see that the low energy SUSY can generally explain the muon $g-2$ at $2\sigma$ level 
and also satisfy other constraints like the 125 GeV Higgs mass, the dark matter 
and the LHC searches. 
We conclude that low energy SUSY is still a compelling candidate for new physics, though suffering
from a little fine-tuning. The future runs of the LHC will continue to explore SUSY 
and could bring surprise at any moment.


\addcontentsline{toc}{section}{Acknowledgments}
\acknowledgments
Thanks to Lei Wu for discussions and comments. 
This work was supported by the National Natural Science Foundation of China 
(NNSFC) under grant Nos. 12105248, 12075213, 11821505, 12075300, 11775012 and 11805161, 
by the Key Research Project of Henan Education Department for colleges and universities 
under grant number 21A140025,  
by Peng-Huan-Wu Theoretical Physics Innovation Center (12047503),
by the CAS Center for Excellence in Particle Physics (CCEPP), 
by the CAS Key Research Program of Frontier Sciences, 
and by a Key R\&D Program of Ministry of Science and Technology of China
under number 2017YFA0402204, by the Key Research Program of the Chinese 
Academy of Sciences, grant No. XDPB15, and by Korea Research Fellowship 
Program through the National Research Foundation of Korea (NRF) funded by the 
Ministry of Science and ICT (2019H1D3A1A01070937).


\begin{thebibliography}{99}
\vspace{-1mm}

 \bibitem{witten} E. Witten, Supersymmetry and Other Scenarios,  
                  talk given at Lepton-Photon Conference, Fermilab, August 16, 2003.
 \bibitem{shifman} M.~Shifman, Musings on the Current Status of HEP, 
                   Mod. Phys. Lett. A35 (2020)2030003
                   [arXiv:2001.00101].
 \bibitem{weinberg} S. Weinberg, Conceptual Foundations of the Unified Theory of Weak and 
                    Electromagnetic Interactions,
                    Nobel Lecture, December 8, 1979. 
 \bibitem{weinberg-2} S. Weinberg, The cosmological constant problem, Rev. Mod. Phys. 61, 1 (1989)
 \bibitem{nima-talk} N. Arkani-Hamed, Naturalness and the muon magnetic momemnt, talk at KIAS, June 11, 2021. 
\bibitem{witten-bj} E. Witten, talk given at International Conference on String Theory, Beijing, Aug 17, 2002
 \bibitem{no-go} S.~R.~Coleman and J.~Mandula, All Possible Symmetries of the S Matrix,
                 Phys. Rev. 159 (1967) 1251.
\bibitem{golfand} Y.~A.~Golfand and E.~P.~Likhtman,
                  Extension of the Algebra of Poincare Group Generators and Violation of p Invariance,
                  JETP Lett. 13 (1971) 323.
\bibitem{haag} R.~Haag, J.~T.~Lopuszanski and M.~Sohnius, 
               All Possible Generators of Supersymmetries of the s Matrix,
                Nucl. Phys. B88 (1975) 257.
\bibitem{ramond} P. Ramond, Dual Theory for Free Fermions, Phys. Rev. D3 (1971) 2415.
\bibitem{neveu}  A. Neveu and J. H. Schwarz, Factorizable dual model of pions, Nucl. Phys. B31 (1971) 86.
\bibitem{sakita}  J. L. Gervais and B. Sakita, Field theory interpretation of supergauges in dual models,
                   Nucl. Phys. B34 (1971) 632.
\bibitem{wess} J. Wess and B. Zumino, 
               Supergauge transformations in four dimensions, Nucl. Phys. B70 (1974) 39.
\bibitem{supergravity} D. Z. Freedman, P. van Nieuwenhuizen, S. Ferrara, 
                       Progress toward a theory of supergravity,
                        Phys. Rev. D13 (1976) 3214.
\bibitem{Martin} S.~P.~Martin, 
                 A Supersymmetry primer, Adv. Ser. Direct. High Energy Phys. 18 (1998) 1
                              [arXiv:hep-ph/9709356 [hep-ph]].
\bibitem{baer-book} H. Baer and X. Tata, Weak Scale Supersymmetry, Cambridge Univeristy Press, 2006.
\bibitem{Gunion:1984yn} J.~F.~Gunion and H.~E.~Haber, Higgs Bosons in Supersymmetric Models. 1.,
                       Nucl. Phys. B \textbf{272} (1986), 1.
\bibitem{Carena:2011aa} M.~Carena, S.~Gori, N.~R.~Shah and C.~E.~M.~Wagner,
   A 125 GeV SM-like Higgs in the MSSM and the $\gamma \gamma$ rate,
   arXiv:1112.3336 [hep-ph].    

\bibitem{HiggsWorkingGroup:2000zik} M.~Carena \textit{et al.} [Higgs Working Group],
                                    Report of the Tevatron Higgs working group,
                                    arXiv:hep-ph/0010338 [hep-ph].

\bibitem{Riesselmann:1997kg} K.~Riesselmann, Limitations of a standard model Higgs boson,
                             arXiv:hep-ph/9711456 [hep-ph].

\bibitem{Hambye:1996wb} T.~Hambye and K.~Riesselmann,
                        Matching conditions and Higgs mass upper bounds revisited,
                        Phys. Rev. D \textbf{55} (1997), 7255
                        [arXiv:hep-ph/9610272 [hep-ph]].

\bibitem{Altarelli:1994rb} G.~Altarelli and G.~Isidori,
                           Lower limit on the Higgs mass in the standard model: An Update,
                           Phys. Lett. B \textbf{337} (1994) 141
\bibitem{Casas:1994qy} J.~A.~Casas, J.~R.~Espinosa and M.~Quiros,
          Improved Higgs mass stability bound in the standard model and implications for supersymmetry,
          Phys. Lett. B \textbf{342} (1995) 171
          [arXiv:hep-ph/9409458 [hep-ph]].
\bibitem{Han:2016xet}
C.~Han, J.~Ren, L.~Wu, J.~M.~Yang and M.~Zhang,
Eur. Phys. J. C \textbf{77} (2017) no.2, 93
doi:10.1140/epjc/s10052-017-4662-7
[arXiv:1609.02361 [hep-ph]].

\bibitem{Kobakhidze:2015scd}
A.~Kobakhidze, N.~Liu, L.~Wu, J.~M.~Yang and M.~Zhang,
Phys. Lett. B \textbf{755} (2016), 76-81
doi:10.1016/j.physletb.2016.02.003
[arXiv:1511.02371 [hep-ph]].

\bibitem{Han:2013kga}
C.~Han, K.~i.~Hikasa, L.~Wu, J.~M.~Yang and Y.~Zhang,
JHEP \textbf{10} (2013), 216
doi:10.1007/JHEP10(2013)216
[arXiv:1308.5307 [hep-ph]].

\bibitem{Cao:2012rz}
J.~Cao, C.~Han, L.~Wu, J.~M.~Yang and Y.~Zhang,
JHEP \textbf{11} (2012), 039
doi:10.1007/JHEP11(2012)039
[arXiv:1206.3865 [hep-ph]].


\bibitem{gluino-cms}
             A.~M.~Sirunyan \textit{et al.} [CMS],
             Search for supersymmetry in pp collisions at $\sqrt{s}=$ 13 TeV with 137 fb$^{-1}$ 
             in final states with a single lepton using the sum of masses of large-radius jets,
             Phys. Rev. D \textbf{101} (2020) 052010
             [arXiv:1911.07558 [hep-ex]].

\bibitem{gluino-atlas} ATLAS Collaboration, 
        Search for squarks and gluinos in final states with jets and 
        missing transverse momentum using 139 fb$^{-1}$ of $\sqrt{s}$ =13 TeV $pp$ 
        collision data with the ATLAS detector,
        ATLAS-CONF-2019-040.

\bibitem{stop-atlas} M.~Aaboud \textit{et al.} [ATLAS],
                   Search for top-squark pair production in final states with one lepton, jets,
                   and missing transverse momentum using 36 fb$^{-1}$ of $\sqrt{s}=13$ TeV pp 
                   collision data with the ATLAS detector,
                   JHEP \textbf{06} (2018) 108
                   [arXiv:1711.11520 [hep-ex]].
\bibitem{stop-cms} 
              A.~M.~Sirunyan \textit{et al.} [CMS],
              Search for direct top squark pair production in events with one lepton, jets, 
              and missing transverse momentum at 13 TeV with the CMS experiment,
              JHEP \textbf{05} (2020) 032
              [arXiv:1912.08887 [hep-ex]].
            
\bibitem{susy-moriond-2021} S. Norberg, Searches for Strong Supersymmetry, 
                Rencontres de Moriond 2021: Electroweak Interactions \& Unified Theories, 
                March 21-27, 2021, http://moriond.in2p3.fr/2021/EW. 

\bibitem{ATLAS:2020pgy} G.~Aad \textit{et al.} [ATLAS],
        Search for direct production of electroweakinos in final states with one lepton, 
        missing transverse momentum and a Higgs boson decaying into two $b$-jets in $pp$ collisions 
        at $\sqrt{s}=13$ TeV with the ATLAS detector,
        Eur. Phys. J. C \textbf{80} (2020) 691
        [arXiv:1909.09226 [hep-ex]].
\bibitem{ATLAS:2019lff} G.~Aad \textit{et al.} [ATLAS],
        Search for electroweak production of charginos and sleptons decaying into final states 
        with two leptons and missing transverse momentum in $\sqrt{s}=13$ TeV $pp$ collisions 
        using the ATLAS detector,
        Eur. Phys. J. C \textbf{80} (2020) 123
        [arXiv:1908.08215 [hep-ex]].
\bibitem{ATLAS:2019wgx} G.~Aad \textit{et al.} [ATLAS],
        Search for chargino-neutralino production with mass splittings near the electroweak 
        scale in three-lepton final states in $\sqrt {s}$=13  TeV $pp$ collisions with the 
        ATLAS detector,
        Phys. Rev. D \textbf{101} (2020) 072001
        [arXiv:1912.08479 [hep-ex]].

\bibitem{baer} H.~Baer, V.~Barger, S.~Salam, D.~Sengupta and K.~Sinha,
          Status of weak scale supersymmetry after LHC Run 2 and ton-scale noble liquid WIMP searches,
          Eur. Phys. J. ST 229 (2020) 3085-3141 [arXiv:2002.03013 [hep-ph]].

\bibitem{Cao:2012fz} J.~J.~Cao, Z.~X.~Heng, J.~M.~Yang, Y.~M.~Zhang and J.~Y.~Zhu,
            A SM-like Higgs near 125 GeV in low energy SUSY: a comparative study for MSSM and NMSSM,
            JHEP \textbf{03} (2012) 086
            [arXiv:1202.5821 [hep-ph]].
\bibitem{higgsino-lsp} C.~Han, A.~Kobakhidze, N.~Liu, A.~Saavedra, L.~Wu and J.~M.~Yang,
                       Probing Light Higgsinos in Natural SUSY from Monojet Signals at the LHC,
                       JHEP 02 (2014) 049 [arXiv:1310.4274 [hep-ph]].
\bibitem{naturalness-1}  C.~Brust, A.~Katz, S.~Lawrence and R.~Sundrum,
                       SUSY, the Third Generation and the LHC,
                       JHEP {\bf 1203}, 103 (2012).
\bibitem{naturalness-2}
                    M.~Papucci, J.~T.~Ruderman and A.~Weiler,
                    Natural SUSY Endures,
                    JHEP {\bf1209}, 035 (2012).
\bibitem{naturalness-3}
                    L.~J.~Hall, D.~Pinner and J.~T.~Ruderman,
                    A Natural SUSY Higgs Near 126 GeV,
                    JHEP {\bf 1204}, 131 (2012).
\bibitem{naturalness-4}
                    J.~L.~Feng and D.~Sanford,
                    A Natural 125 GeV Higgs Boson in the MSSM from Focus Point Supersymmetry with A-Terms,
                    Phys.\ Rev.\ D {\bf86}, 055015 (2012).

\bibitem{Baer:2012up} H.~Baer, V.~Barger, P.~Huang, A.~Mustafayev and X.~Tata,
                      Radiative natural SUSY with a 125 GeV Higgs boson,
                      Phys.\ Rev.\ Lett.\  {\bf 109}, 161802 (2012).
\bibitem{Tata:2020afe} X.~Tata, Natural supersymmetry: status and prospects,
                       Eur. Phys. J. ST \textbf{229} (2020) no.21, 3061-3083
                       [arXiv:2002.04429 [hep-ph]].
\bibitem{GAMBIT:2018gjo}
P.~Athron \textit{et al.} [GAMBIT],
Combined collider constraints on neutralinos and charginos,
Eur. Phys. J. C \textbf{79} (2019) no.5, 395
[arXiv:1809.02097 [hep-ph]].

\bibitem{stop-lsp-1} G.~H.~Duan, X.~Fan, K.~Hikasa, B.~Peng and J.~M.~Yang,
                   Probing stops in the coannihilation region at the HL-LHC: a comparative study 
                   of different processes,
                   Phys. Lett. B810 (2020) 135800 [arXiv:1912.01970 [hep-ph]]. 
\bibitem{stop-lsp-2}
                   M.~Abdughani, J.~Ren, L.~Wu and J.~M.~Yang,
                   Probing stop pair production at the LHC with graph neural networks,
                   JHEP 08 (2019) 055 [arXiv:1807.09088 [hep-ph]].
\bibitem{bino-wino} G.~H.~Duan, K.~Hikasa, J.~Ren, L.~Wu and J.~M.~Yang,
                    Probing bino-wino coannihilation dark matter below the neutrino floor at the LHC,
                    Phys. Rev. D98 (2018) 015010 [arXiv:1804.05238 [hep-ph]].
\bibitem{mulat-bino-higgsino} M.~Abdughani, L.~Wu and J.~M.~Yang,
                          Status and prospects of light bino-higgsino dark matter in natural SUSY,
                          Eur. Phys. J. C78 (2018)4 [arXiv:1705.09164 [hep-ph]].
\bibitem{gm2-mssm} M.~Abdughani, K.~I.~Hikasa, L.~Wu, J.~M.~Yang and J.~Zhao,
                  Testing electroweak SUSY for muon g-2 and dark matter at the LHC and beyond,
                 JHEP 11 (2019) 095, arXiv:1909.07792. 
\bibitem{gm2-mssm-2} P.~Cox, C.~Han and T.~T.~Yanagida,
                 Muon g-2 and dark matter in the minimal supersymmetric standard model,
                 Phys. Rev. D 98 (2018) 055015, arXiv:1805.02802.
\bibitem{gm2-mssm-3}
              P.~Athron, C.~Bal\'azs, D.~H.~Jacob, W.~Kotlarski, D.~St\"ockinger and H.~St\"ockinger-Kim,
              New physics explanations of $a_\mu$ in light of the FNAL muon $g-2$ measurement,
              arXiv:2104.03691.
\bibitem{gm2-mssm-4}
        H.~Baer, V.~Barger and H.~Serce,
        Anomalous muon magnetic moment, supersymmetry, naturalness, LHC search limits and the landscape,
        arXiv:2104.07597.
\bibitem{gm2-mssm-5}
        W.~Altmannshofer, S.~A.~Gadam, S.~Gori and N.~Hamer,
        Explaining g-2 with Multi-TeV Sleptons, arXiv:2104.08293.
\bibitem{gm2-mssm-6}
        M. Chakraborti, S.v Heinemeyer, I. Saha, C. Schappacher, $(g-2)_\mu$ and SUSY Dark Matter:
        Direct Detection and Collider Search Complementarity, arXiv:2112.01389.
\bibitem{gm2-mssm-7}
        W.~Ahmed, I.~Khan, J.~Li, T.~Li, S.~Raza and W.~Zhang,
        The Natural Explanation of the Muon Anomalous Magnetic Moment via the Electroweak 
        Supersymmetry from the GmSUGRA in the MSSM, arXiv:2104.03491 [hep-ph].
\bibitem{Pozzo:2018anw} G.~Pozzo and Y.~Zhang,
        Constraining resonant dark matter with combined LHC electroweakino searches,
        Phys. Lett. B \textbf{789} (2019), 582-591
        [arXiv:1807.01476 [hep-ph]].
\bibitem{fit-cmssm} C.~Han, K.~Hikasa, L.~Wu, J.~M.~Yang and Y.~Zhang,
                    Status of CMSSM in light of current LHC Run-2 and LUX data,
                   Phys. Lett. B 769 (2017) 470 [arXiv:1612.02296 [hep-ph]];
\bibitem{fit-cmssm-2} 
P.~Athron \textit{et al.} [GAMBIT], Global fits of GUT-scale SUSY models with GAMBIT,
Eur. Phys. J. C \textbf{77} (2017) 824
[arXiv:1705.07935 [hep-ph]].


\bibitem{Rinchiuso:2020skh}
L.~Rinchiuso, O.~Macias, E.~Moulin, N.~L.~Rodd and T.~R.~Slatyer,
Prospects for detecting heavy WIMP dark matter with the Cherenkov Telescope Array: The Wino and Higgsino,
Phys. Rev. D \textbf{103} (2021) 023011
[arXiv:2008.00692 [astro-ph.HE]].

\bibitem{marco} M. Cirelli, A. Strumia, and M. Tamburini, 
Cosmology and Astrophysics of Minimal Dark Matter, Nucl. Phys. B787 (2007) 152,
arXiv:0706.4071 [hep-ph].
\bibitem{hisano} J. Hisano, S. Matsumoto, M. Nagai, O. Saito, and M. Senami, 
Non-perturbative effect on thermal relic abundance of dark matter, 
Phys. Lett. B646 (2007) 34, arXiv: 0610249 [hep-ph].
\bibitem{wino-lsp2}  A. Hryczuk, R. Iengo, and P. Ullio,
                    Relic densities including Sommerfeld enhancements in the MSSM, JHEP 03 (2011) 069, 
                    arXiv:1010.2172 [hep-ph].
\bibitem{wino-lsp3} M. Beneke, A. Bharucha, F. Dighera, C. Hellmann, A. Hryczuk, S. Recksiegel, and P. Ruiz-Femenia, 
                    Relic density of wino-like dark matter in the MSSM, 
                    JHEP 03 (2016) 119, arXiv:1601.04718 [hep-ph].


\bibitem{Feng:2004mt} J.~L.~Feng, S.~Su and F.~Takayama,
                     Supergravity with a gravitino LSP, Phys. Rev. D \textbf{70} (2004) 075019
                     [arXiv:hep-ph/0404231 [hep-ph]].
\bibitem{Wang:2004ib} F.~Wang and J.~M.~Yang,
                     SuperWIMP dark matter scenario in light of WMAP,
                     Eur. Phys. J. C \textbf{38} (2004) 129 [arXiv:hep-ph/0405186 [hep-ph]].
\bibitem{Gu:2020ozv} Y.~Gu, M.~Khlopov, L.~Wu, J.~M.~Yang and B.~Zhu,
                     Light gravitino dark matter: LHC searches and the Hubble tension,
                     Phys. Rev. D \textbf{102} (2020) 115005
                     [arXiv:2006.09906 [hep-ph]].
\bibitem{Argurio:2011hs} R.~Argurio, Z.~Komargodski and A.~Mariotti,
                        Pseudo-Goldstini in Field Theory,
                        Phys. Rev. Lett. \textbf{107} (2011) 061601
                        [arXiv:1102.2386 [hep-th]].
\bibitem{Dai:2021eah} J.~Dai, T.~Liu and J.~M.~Yang,
             An explicit calculation of pseudo-goldstino mass at the leading three-loop level,
             JHEP \textbf{06} (2021) 175
             [arXiv:2104.12656 [hep-ph]].
\bibitem{Moroi:1993mb} T.~Moroi, H.~Murayama and M.~Yamaguchi,
                      Cosmological constraints on the light stable gravitino,
                      Phys. Lett. B \textbf{303} (1993) 289.
\bibitem{Wu:gravitino} Y.~Gu, M.~Khlopov, L.~Wu, J.~M.~Yang and B.~Zhu,
Phys. Rev. D \textbf{102}, no.11, 115005 (2020)
doi:10.1103/PhysRevD.102.115005
[arXiv:2006.09906 [hep-ph]].
\bibitem{Fei:2007.09981} J.~Cao, X.~Du, Z.~Li, F.~Wang and Y.~Zhang,
[arXiv:2007.09981 [hep-ph]].
                      
\bibitem{Liu:2013sx} T.~Liu, L.~Wang and J.~M.~Yang,
                     Higgs decay to goldstini and its observability at the LHC,
                     Phys. Lett. B \textbf{726} (2013) 228
                     [arXiv:1301.5479 [hep-ph]].
\bibitem{Hikasa:2014yra} K.~Hikasa, T.~Liu, L.~Wang and J.~M.~Yang,
                         Pseudo-goldstino and electroweak gauginos at the LHC,
                         JHEP \textbf{07} (2014) 065
                         [arXiv:1403.5731 [hep-ph]].
\bibitem{Liu:2014lda} T.~Liu, L.~Wang and J.~M.~Yang,
                      Pseudo-goldstino and electroweakinos via VBF processes at LHC,
                      JHEP \textbf{02} (2015) 177
                      [arXiv:1411.6105 [hep-ph]].
\bibitem{Franzosi:2021zwp} D.~B.~Franzosi, G.~Ferretti, E.~Riefel and S.~Strandberg,
            Electroweak signatures of gauge-mediated supersymmetry breaking in multiple hidden sectors,
            [arXiv:2111.04775 [hep-ph]].
\bibitem{Chen:2021omv} J.~Chen, C.~Han, J.~M.~Yang and M.~Zhang,
                       Probing a bino NLSP at lepton colliders,
                       Phys. Rev. D \textbf{104} (2021) 015009
                       [arXiv:2101.12131 [hep-ph]].
\bibitem{FNAL:gmuon}  Muon g-2 Collaboration, Measurement of the positive muon anomalous magnetic moment to 0.46 ppm, 
                      Phys. Rev. Lett. 126 (2021) 141801.
 \bibitem{BNL:gmuon} Muon g-2 Collaboration, Final Report of the Muon E821 Anomalous Magnetic Moment Measurement at BNL,
                     Phys. Rev. D 73  (2006) 072003.
\bibitem{sm-prediction-1} T. Aoyama, et al., The anomalous magnetic moment of the muon in the standard model, 
                        Phys. Rep. 887 (2020) 1.
\bibitem{sm-prediction-2}
                    M. Davier, A. Hoecker, B. Malaescu, Z. Zhang, 
                    A new evaluation of the hadronic vacuum polarisation 
                    contribu-tions to the muon anomalous magnetic moment and to $\alpha(m^2_Z)$, 
                    Eur. Phys. J. C 80 (2020) 241.
\bibitem{gm2-lattice} S.~Borsanyi, et al.,  Leading hadronic contribution to the muon magnetic moment from lattice QCD,
                      Nature 593 (2021) 51.  
\bibitem{moroi}  T. Moroi, 
                 The muon anomalous magnetic dipole moment in the minimal supersymmetric standard model, 
                 Phys. Rev. D 53 (1996) 6565.
\bibitem{stockinger} D. Stockinger, The muon magnetic moment and supersymmetry, J. Phys. G 34 (2007) R45.
\bibitem{gm2-gutsusy} F.~Wang, L.~Wu, Y.~Xiao, J.~M.~Yang and Y.~Zhang,
                      GUT-scale constrained SUSY in light of E989 muon g-2 measurement,
                      arXiv:2104.03262.

\bibitem{gm2-gutsusy-2}                      
                 M.~Chakraborti, L.~Roszkowski and S.~Trojanowski,
                 GUT-constrained supersymmetry and dark matter in light of the new $(g-2)_\mu$ determination,
                      JHEP 05 (2021) 252, arXiv:2104.04458.
\bibitem{gm2-gutsusy-3}   
                      A.~Aboubrahim, P.~Nath and R.~M.~Syed,
                      Yukawa coupling unification in an SO(10) model consistent with Fermilab 
                      g-2 result,
                      JHEP 06 (2021) 002, arXiv:2104.10114.
                    
\bibitem{GMSB-yukawa-higgs-1} Z. Kang, T. Li, T. Liu, C. Tong, J.M. Yang, 
                            A heavy SM-like Higgs and a light stop from Yukawa-deflected gauge mediation, 
                            Phys. Rev. D 86 (2012) 095020 [arXiv: 1203.2336 [hep-ph]].
\bibitem{GMSB-yukawa-higgs-2} 
                            J.~L.~Evans, M.~Ibe, S.~Shirai and T.~T.~Yanagida,
                            A 125GeV Higgs Boson and Muon g-2 in More Generic Gauge Mediation,
                            Phys. Rev. D \textbf{85} (2012), 095004
                            [arXiv:1201.2611 [hep-ph]].
\bibitem{gm2-SUGRA-ext-1} S.~Akula and P.~Nath,
            Gluino-driven radiative breaking, Higgs boson mass, muon g-2, and the Higgs diphoton decay in 
            supergravity unification, 
                        Phys. Rev. D \textbf{87}, 115022  (2013), arXiv:1304.5526.
\bibitem{gm2-SUGRA-ext-2}
                       Z.~Li, G.~L.~Liu, F.~Wang, J.~M.~Yang and Y.~Zhang,
                       Gluino-SUGRA scenarios in light of FNAL muon g-2 anomaly, arXiv:2106.04466;\\
\bibitem{gm2-SUGRA-ext-3} F.~Wang, W.~Wang and J.~M.~Yang,
                       Reconcile muon g-2 anomaly with LHC data in SUGRA with generalized gravity mediation,
                       JHEP 06, 079 (2015), arXiv:1504.00505.
\bibitem{gm2-SUGRA-ext-4}  F.~Wang, K.~Wang, J.~M.~Yang and J.~Zhu,
                       Solving the muon g-2 anomaly in CMSSM extension with non-universal gaugino masses,
                       JHEP 12, 041 (2018), arXiv:1808.10851.
\bibitem{gm2-SUGRA-ext-5}  F. Wang, W. Wang, J.M. Yang, Y. Zhang, 
                       Heavy colored SUSY partners from deflected anomaly mediation, 
                       JHEP 07 (2015) 138, arXiv :1505.02785.
\bibitem{gm2-SUGRA-ext-6}
             F. Wang, J. M. Yang, Y. Zhang, 
             Radiative natural SUSY spectrum from deflected AMSB scenario with messenger-matter interactions, 
                       JHEP 04 (2016) 177, arXiv :1602.01699.
  \bibitem{super:GUT}  J.R. Ellis, A. Mustafayev and K.A. Olive, Eur. Phys. J. C 69 (2010) 201 [arXiv:1003.3677] .
\bibitem{sub:GUT} J.R. Ellis, K.A. Olive and P. Sandick, Phys. Lett. B 642 (2006) 389 [hep-ph/0607002] [INSPIRE]                 
 \bibitem{Fei:1508.01299} F.~Wang,
Phys. Lett. B \textbf{751}, 402-407 (2015) [arXiv:1508.01299 [hep-ph]].
 \bibitem{Fei:1512.06715} F.~Wang, L.~Wu, J.~M.~Yang and M.~Zhang,
Phys. Lett. B \textbf{759}, 191-199 (2016)
[arXiv:1512.06715 [hep-ph]].
 \bibitem{Fei:1703.10894} F.~Wang, W.~Wang and J.~M.~Yang,
Phys. Rev. D \textbf{96}, no.7, 075025 (2017)
[arXiv:1703.10894 [hep-ph]].
 \bibitem{Fei:1710.06105} X.~Du and F.~Wang,
Eur. Phys. J. C \textbf{78}, no.5, 431 (2018)
[arXiv:1710.06105 [hep-ph]].
 \bibitem{Fei:1903.05669}    Z.~Li and F.~Wang,
JHEP \textbf{05}, 009 (2019)
[arXiv:1903.05669 [hep-ph]].                 
   \bibitem{Fei:1804.07335} X.~K.~Du, G.~L.~Liu, F.~Wang, W.~Wang, J.~M.~Yang and Y.~Zhang,
Eur. Phys. J. C \textbf{79}, no.5, 397 (2019)
[arXiv:1804.07335 [hep-ph]].
   \bibitem{Fei:1808.08529}  F.~Wang,
JHEP \textbf{11}, 062 (2018)
[arXiv:1808.08529 [hep-ph]].                  
\bibitem{gm2-nmssm-1}  X. Ning, F. Wang, 
                     Solving the muon g-2 anomaly within the NMSSM from generalized deflected AMSB, 
                     JHEP 08 (2017) 089, arXiv: 1704.05079.
\bibitem{gm2-nmssm-2}
                     M. Abdughani, Y.-Z. Fan, L. Feng, Y.-L. Tsai, L. Wu and Q. Yuan, 
                     A common origin of muon g-2 anomaly, Galaxy Center GeV excess and AMS-02 anti-proton 
                     excess in the NMSSM, arXiv:2104.03274.
\bibitem{gm2-nmssm-3}
                     J.~Cao, J.~Lian, Y.~Pan, D.~Zhang and P.~Zhu,
                     Improved $(g-2)_\mu$ Measurement and Singlino dark matter in the general NMSSM,
                     arXiv:2104.03284.

\bibitem{Wang:2021lwi} K.~Wang and J.~Zhu,
A smuon in the NMSSM confronted with the muon g-2 and SUSY searches,
arXiv:2112.14576 [hep-ph].

\bibitem{berkeley} R. H. Parker, C. Yu, W. Zhong, B. Estey and H. Muller, 
                   Measurement of the fine-structure constant as a test of the Standard Model, Science 360 (2018) 191.

\bibitem{joint-1}  S.~Li, Y.~Xiao and J.~M.~Yang,
                 Can electron and muon $g-2$ anomalies be jointly explained in SUSY ?, arXiv:2107.04962.
\bibitem{joint-2} 
                 B. Dutta and Y. Mimura, Electron g-2 with flavor violation in MSSM, 
                 Phys. Lett. B 790 (2019) 563.
\bibitem{joint-3} 
         M. Badziak and K. Sakurai, 
         Explanation of electron and muon g-2 anomalies in the MSSM, JHEP 10 (2019) 024.
\bibitem{joint-4} 
         M. Endo and W. Yin, Explaining electron and muon g-2 anomaly in SUSY without lepton-flavor mixings, 
         JHEP 08 (2019) 122.
\bibitem{joint-5} 
         J. Cao, Y. He, J. Lian, D. Zhang and P. Zhu, Electron and Muon Anomalous Magnetic
         Moments in the Inverse Seesaw Extended NMSSM, arXiv:2102.11355[hep-ph].
\bibitem{Ali:2021kxa}
        M.~I.~Ali, M.~Chakraborti, U.~Chattopadhyay and S.~Mukherjee,
        Muon and Electron $(g-2)$ Anomalies with Non-Holomorphic Interactions in MSSM,
        arXiv:2112.09867 [hep-ph].

\bibitem{cp-phase-1} C. Han, Muon g-2 and CP violation in MSSM, arXiv:2104.03292 [hep-ph].
\bibitem{cp-phase-2} S.~Li, Y.~Xiao and J.~M.~Yang,
         Constraining CP-phases in SUSY: An interplay of muon/electron g-2 and electron EDM,
         Nucl. Phys. B \textbf{974} (2022) 115629
         arXiv:2108.00359 [hep-ph].

\bibitem{b-decay-susy} M.~A.~Boussejra and F.~Mahmoudi,
                       New constraints on flavour violating supersymmetry, arXiv:2111.07938 [hep-ph].

\bibitem{Papucci:2011wy} M.~Papucci, J.~T.~Ruderman and A.~Weiler, Natural SUSY Endures,
                         JHEP \textbf{09} (2012), 035
                         [arXiv:1110.6926 [hep-ph]].
\bibitem{Baer:2016bwh} H.~Baer, V.~Barger, N.~Nagata and M.~Savoy,
                 Phenomenological profile of top squarks from natural supersymmetry at the LHC,
                 Phys. Rev. D \textbf{95} (2017)  055012
                 [arXiv:1611.08511 [hep-ph]].

\bibitem{Barbieri:1987fn}  R.~Barbieri and G.~F.~Giudice,
                    Upper Bounds on Supersymmetric Particle Masses,
                    Nucl. Phys. B \textbf{306} (1988) 63.

\bibitem{Ellis:1986yg} J.~R.~Ellis, K.~Enqvist, D.~V.~Nanopoulos and F.~Zwirner,
                    Observables in Low-Energy Superstring Models,
                    Mod. Phys. Lett. A \textbf{1} (1986) 57.

\bibitem{Baer:2013gva} H.~Baer, V.~Barger and D.~Mickelson,
                       How conventional measures overestimate electroweak fine-tuning in supersymmetric theory,
                       Phys. Rev. D \textbf{88} (2013) 095013
                       [arXiv:1309.2984 [hep-ph]].

\bibitem{Baer:2012cf} H.~Baer, V.~Barger, P.~Huang, D.~Mickelson, A.~Mustafayev and X.~Tata,
        Radiative natural supersymmetry: Reconciling electroweak fine-tuning and the Higgs boson mass,
        Phys. Rev. D \textbf{87} (2013) 115028
        [arXiv:1212.2655 [hep-ph]].

\bibitem{Cohen:2020ohi} T.~Cohen, N.~Craig, S.~Koren, M.~Mccullough and J.~Tooby-Smith,
                        Supersoft Top Squarks,
                        Phys. Rev. Lett. \textbf{125} (2020) 151801
                        [arXiv:2002.12630 [hep-ph]].

\bibitem{Li:2020glc} H.~Li, H.~Song, S.~Su, W.~Su and J.~M.~Yang, 
                     MSSM at future Higgs factories,
                     Chin. Phys. C \textbf{45} (2021) 045106
                     [arXiv:2010.09782 [hep-ph]].

\bibitem{Huang:2014ifa} W.~Huang, Z.~Kang, J.~Shu, P.~Wu and J.~M.~Yang,
                        New insights in the electroweak phase transition in the NMSSM,
                        Phys. Rev. D \textbf{91} (2015) no.2, 025006
                        [arXiv:1405.1152 [hep-ph]].
\bibitem{split-1} N. Arkani-Hamed and S. Dimopoulos, 
  Supersymmetric unification without low energy supersymmetry and signatures for fine-tuning at the LHC, 
   JHEP 06 (2005) 073 [hep-th/0405159].
\bibitem{split-2} G. F. Giudice and A. Romanino, Split supersymmetry, 
                  Nucl. Phys. B 699 (2004) 65 [hep-ph/0406088].
\bibitem{split-3} N. Arkani-Hamed, S. Dimopoulos, G. F. Giudice and A. Romanino, 
     Aspects of split supersymmetry, Nucl. Phys. B 709 (2005) 3 [hep-ph/0409232].
\bibitem{Wang:2015mea} F.~Wang, W.~Wang and J.~M.~Yang,
                       A split SUSY model from SUSY GUT,
                      JHEP \textbf{03} (2015) 050
                     [arXiv:1501.02906 [hep-ph]].

\bibitem{witten-reflect} E. Witten, Witten reflects, CERN Courier, 21 December, 2021, https://cerncourier.com/a/witten-reflects/ . 

\bibitem{Wang:2013rba} F.~Wang, W.~Wang and J.~M.~Yang,
Split supersymmetry under GUT and current dark matter constraints,
Eur. Phys. J. C \textbf{74} (2014) 3121
[arXiv:1310.1750 [hep-ph]].

\bibitem{Wang:2005kf} F.~Wang, W.~Wang and J.~M.~Yang,
   Dark matter constraints on gaugino/Higgsino masses in split supersymmetry and their implications at colliders,
  Eur. Phys. J. C \textbf{46} (2006) 521
  [arXiv:hep-ph/0512133 [hep-ph]].


\end{thebibliography}
\end{document}